\documentclass[12pt]{article}
\pdfoutput=1

\usepackage[square,comma,numbers,sort&compress]{natbib}
\usepackage[left=20mm,right=20mm,top=20mm,bottom=20mm]{geometry}
\usepackage{mediabb}
\usepackage{graphics}
\usepackage{epsf}
\usepackage{color}
\usepackage{amsmath}
\usepackage{amssymb}
\usepackage{latexsym}
\usepackage{slashed}
\usepackage{float}
\usepackage{cases}
\usepackage{bm}

\newcommand{\vev}[1]{\left\langle#1\right\rangle}

\newcommand{\D}{\mathcal{D}}
\newcommand{\E}{\mathcal{E}}

\renewcommand{\H}{\mathcal{H}}

\newcommand{\U}{\mathcal{U}}

\newcommand{\Y}{\mathcal{Y}}

\newcommand{\al}[1]{\begin{align}#1\end{align}}

\newcommand{\bp}{\begin{pmatrix}}
\newcommand{\ep}{\end{pmatrix}}
\newcommand{\bb}{\begin{bmatrix}}
\newcommand{\eb}{\end{bmatrix}}

\newcommand{\paren}[1]{\left(#1\right)}
\newcommand{\sqbr}[1]{\left[#1\right]}
\newcommand{\ab}[1]{\left|#1\right|}

\newcommand{\br}[1]{\left\{#1\right\}}

\newcommand{\beq}{\begin{equation}}
\newcommand{\eeq}{\end{equation}}
\newcommand{\bea}{\begin{eqnarray}}
\newcommand{\eea}{\end{eqnarray}}

\newcommand{\dis}{\displaystyle}

\newcommand{\pal}{\partial}

\newcommand{\fulltoday}{\number\day\space \ifcase\month\or
    January\or February\or March\or April\or May\or June\or
    July\or August\or September\or October\or November\or December\fi
    \space\number\year}

 \makeatletter
    \renewcommand{\theequation}{%
    \thesection.\arabic{equation}}
    \@addtoreset{equation}{section}
  \makeatother

\begin{document}
%\maketitle
\begin{titlepage}
\renewcommand\thefootnote{\alph{footnote}}
		\mbox{}\hfill KOBE-TH-13-1\\
		\mbox{}\hfill HRI-P-13-01-001\\
		\mbox{}\hfill RECAPP-HRI-2013-001\\
\vspace{4mm}
\begin{center}
{\fontsize{22pt}{0pt}\selectfont \bf
{$CP$ phase from twisted Higgs\\ vacuum expectation value
in extra dimension}} \\
\vspace{8mm}
	{\fontsize{13pt}{0pt}\selectfont \bf
	Yukihiro Fujimoto,\,\footnote{
		E-mail: \tt {{fujimoto@crystal.kobe-u.ac.jp}}
		}
	{}
	Kenji Nishiwaki,\,\footnote{
		E-mail: \tt nishiwaki@hri.res.in
		}
	{} and
	Makoto Sakamoto\,\,\footnote{
		E-mail: \tt dragon@kobe-u.ac.jp} } \\
\vspace{8mm}
%	\medskip\\
	{\fontsize{13pt}{0pt}\selectfont
		${}^{\mathrm{a,c}}$\it Department of Physics, Kobe University, Kobe 657-8501, Japan \smallskip\\
		${}^{\mathrm{b}}$\it Regional Centre for Accelerator-based Particle Physics, \\[2pt]
		\it Harish-Chandra Research Institute, Allahabad 211 019, India
		\smallskip\\
	}
\vspace{4mm}
{\normalsize \fulltoday}
\vspace{10mm}
\end{center}
\begin{abstract}
{\fontsize{12pt}{16pt}\selectfont{
We propose a new mechanism for generating {a $CP$ phase via Higgs a} vacuum expectation value originating from geometry of an extra dimension. A twisted boundary condition is the key to produce an {extra-dimension coordinate-dependent} vacuum expectation value, which contains a {$CP$} phase degree of freedom and can be a new source of {a $CP$ phase in higher-dimensional} gauge theories. As an illustrative example, we apply our mechanism to a five-dimensional gauge theory with point interactions and show that our mechanism can dynamically produce a {nontrivial $CP$-violating} phase {with electroweak symmetry breaking}, even though the five-dimensional model does not include any {$CP$-violating} phases of Yukawa couplings in the five-dimensional Lagrangian because of a single generation of five-dimensional fermions.
{We apply our mechanism to a model with point interactions, which has no
source of {$CP$-violating} phases in the couplings of the {higher-dimensional} action, and show that a {nontrivial $CP$} phase dynamically appears.}
}}
\end{abstract}
%\vfill
%		\mbox{}\hfill KOBE-TH-??\\
%		\mbox{}\hfill HRI-P-12-??-???\\
%		\mbox{}\hfill RECAPP-HRI-2012-???
\end{titlepage}
\renewcommand\thefootnote{\arabic{footnote}}
\setcounter{footnote}{0}
%\newpage

%%%%%%
%\begin{itembox}{History}
%\begin{itemize}
%\item {\bf Ver. 0.80:} First published version.
%\end{itemize}
%\end{itembox}
%%%%%%

%\tableofcontents

%%%%%%%%%%%%%%%%%%%%%%%%%%%%%%%%%%%%%%%%%%%%%
%%%%%%%%%%%%%%%%%%%%%%%%
\section{Introduction}
%%%%%%%%%%%%%%%%%%%%%%%%
%%%%%%%%%%%%%%%%%%%%%%%%%%%%%%%%%%%%%%%%%%%%%
Pursuing the origin of the generations of the fermions is one of the important {themes} in particle physics. The three generations (or more) are necessary to produce the Kobayashi--Maskawa {$CP$} phase, which causes {$CP$-violating} effects and was proposed in Ref.~\cite{Kobayashi:1973fv}. If the number of generations were less than {$3$}, any complex phases in the Cabbibo--Kobayashi--Maskawa (CKM) matrix would be absorbed into phases of quark fields and then the Kobayashi-Maskawa {$CP$-violating} mechanism would not work.

Extra-dimensional field theory is one of {the} appealing candidates beyond the standard model {{(SM)}}.
{Many studies have been done {up to} today based on many ideas for pursuing the origin of fermion flavor~\cite{ArkaniHamed:1998vp,ArkaniHamed:1999dc,Dvali:1999cn,Yoshioka:1999ds,Mohapatra:1999zd,Grossman:1999ra,Dvali:2000ha,Gherghetta:2000qt,Huber:2000ie,Libanov:2000uf,Frere:2000dc,Neronov:2001qv,Frere:2001ug,Kaplan:2001ga,Parameswaran:2006db,Gogberashvili:2007gg,Kaplan:2011vz}.}
Especially, we can find several attractive models to solve the generation problem, in which the three generations of the four-dimensional chiral fermions are dynamically realized from a single generation of higher-dimensional fermions. However, such models have a common problem: The number of higher-dimensional Yukawa couplings is not enough to produce a {$CP$-violating} phase because of a single generation of fermions in higher-dimensions, {so} we could not obtain a {$CP$-violating} phase {\it{\`a} la} Kobayashi-Maskawa. Therefore, in models to solve the generation problem, we need some new sources of {$CP$-violating} phases other than higher-dimensional Yukawa couplings. Otherwise, those models without a {$CP$-violating} phase should be discarded as phenomenological ones.\footnote{
In the gauge-Higgs unification model, a similar problem arises because of lack of degree of freedom in the Yukawa sector of an original {five-dimensional} action.
The ways to overcome this point have been studied~\cite{Burdman:2002se,Adachi:2012zk,Lim:2009pj}.}

In this {paper}, we propose a new mechanism to produce a {$CP$} phase in the context of five-dimensional gauge theories. Allowing a twisted boundary condition (BC) for the Higgs doublet leads to a Higgs vacuum expectation value (VEV) with an {extra-dimension coordinate-dependent} phase, which contains a {$CP$} phase degree of freedom. The properties of {such kinds of scalar VEVs} have been studied in {Refs.}~\cite{Sakamoto:1999yk,Sakamoto:1999ym, Sakamoto:1999iv,Ohnishi:2000hs,Hatanaka:2000zq,Matsumoto:2001fp,Sakamoto:2001gn,Coradeschi:2007gb,Burgess:2008ka,Haba:2009uu}.\footnote{
Point interactions on $S^{1}$, which are additional boundary points
(on $S^1$), have been studied in Refs~\cite{Hatanaka:1999ac,Nagasawa:2002un,Nagasawa:2003tw,Nagasawa:2005kv,Nagasawa:2008an}.
We can consider another possibility that some terms are localized in boundary points at tree level~\cite{Flacke:2008ne,Datta:2012xy,Datta:2012tv,Flacke:2012ke}.
}
{We note that the electroweak symmetry is dynamically broken at that time.}

As a demonstration of our mechanism, we apply the mechanism to a five-dimensional gauge theory with point interactions in which three generations in four dimensions are produced from a single generation in five dimensions. We show that a {nontrivial CP} phase dynamically appears in the CKM matrix in four dimensions, even though any coupling constants in the five-dimensional Lagrangian have no {$CP$} phases. Our purpose of this paper is to show that our mechanism does work as a new source of the {$CP$} violation.

This {paper} is organized as follows{.}
In {Sec.}~2, we discuss and verify a possibility of the Higgs doublet with a twisted BC to explain the origin of the {$CP$} phase in the CKM matrix.
In {Sec.}~3, we construct a model with point interactions and a scalar singlets
{for which the} VEV depends on the extra coordinate exponentially.
In {Sec.}~4, we check that the CP phase originating from our mechanism can explain the CKM properties in the above model.
Here, we also discuss the properties of the realized quark masses and other mixings briefly.
In {Sec.}~5, we summarize our results and discuss
some aspects of our model.
In the {Appendix}, details of choosing parameters is explored.

%%%%%%%%%%%%%%%%%%%%%%%%%%%%%%%%%%%%%%%%%%%%%
%%%%%%%%%%%%%%%%%%%%%%%%
\section{Position-dependent VEV (also as {$CP$} phase) with twisted boundary condition
\label{section_2}}
%%%%%%%%%%%%%%%%%%%%%%%%
%%%%%%%%%%%%%%%%%%%%%%%%%%%%%%%%%%%%%%%%%%%%%

In this section, we propose a new mechanism for generating {$CP$} phase with twisted boundary condition of a {five-dimensional} scalar $H$ on $S^1$.
Hereafter, we use a coordinate $y$ to indicate the position in the extra space.
A key aspect is that broken phase can be realized with the scalar{,} and at the same time, the VEV profile itself turns out to be $y$-position dependent and complex, which means that the scalar VEV possibly triggers the {$CP$} violation.
Interestingly, the $y$-position dependence disappears in the gauge boson masses,
even though the VEV of $H$ depends on $y$. This is because the {$y$ dependence} of
the VEV of $H$ is cancelled out in the squared form $H^{\dagger} H$.
This property is very important and it works as an usual {four-dimensional} Higgs mechanism
without violating electroweak precision measurements at the tree level.
When we consider the situation that $H$ is the $SU(2)_W$ Higgs doublet, we can dynamically generate both the suitable electroweak symmetry breaking (EWSB) and the {$CP$-violating} phase simultaneously.
We note that its $SU(N)$ extension is possible and straightforward.

The action we consider is
\al{
S_H &= \int d^4x \int_0^L dy \bigg\{ {H^{\dagger} (\pal_M \pal^M  + {M}^2)  H - \frac{\lambda}{2} (H^{\dagger} H)^2} \bigg\},
\label{doubletaction}
}
where {$M$ and $\lambda$}
are the bulk mass and quartic coupling, 
respectively.
Since $S^{1}$ is a {multiply connected} space, we can impose the 
twisted boundary condition on $H$ as \cite{Sakamoto:1999yk,Sakamoto:1999ym, Sakamoto:1999iv,Ohnishi:2000hs,Hatanaka:2000zq}
\al{
H(y+L) = e^{i \theta} H(y).
\label{twistBC}
}
Here, we take the range of $\theta$ as $-\pi < \theta \leq \pi$.
$L$ shows the circumference of $S^1$, and
we choose the metric convention as
$
\eta_{MN} = \eta^{MN} = \text{diag}(-1,1,1,1,1).
$
The {Latin} indices run from $0$ to {{$3,\,5\ (\text{or\ }y)$,}} and {Greek} ones
run from $0$ to $3$, respectively.

{We note that the VEV of $\langle H(y) \rangle$ should be determined by
minimizing the functional}
\al{
\mathcal{E}[H] = \int_0^{L} dy \bigg\{ {\ab{\pal_y H}^2 - {M}^2 \ab{H}^2 + \frac{\lambda}{2} \ab{H}^4} \bigg\}{,}
}
because the VEV 
can possess the {$y$ dependence}
to minimize the energy.
Here, we assume that the four-dimensional (4D) Lorentz invariance is unbroken.

After introducing $\H(y)$ by
\al{
{H(y) = e^{i {\frac{\theta}{L} y}} \H(y), \quad \H(y+L) = \H(y),}
\label{modified_H}
}
the functional $\E$ can be rewritten as
\al{
\E[H] &= \E_{1}[\H] + \E_{2}[\H], \\
\E_{1}[\H] &= \int_0^L dy \bigg\{
   {\ab{\pal_y \H}^2} + i \frac{\theta}{L} \Big( (\pal_y \H)^{\dagger} \H - \H^{\dagger} \pal_y \H \Big)
   \bigg\}, \\
\E_{2}[\H] &= \int_0^L dy \bigg\{ { \frac{\lambda}{2} \left( \ab{\H}^2 
   - \frac{1}{\lambda}
   \left[ {M}^2 - \left(\frac{\theta}{L}\right)^2 \right]\right)^2
   - \frac{1}{2\lambda} \left[ {M}^2 
   - \left(\frac{\theta}{L}\right)^2\right]^2}
   \bigg\},
}
where ${\cal E}_1$ corresponds to the contribution from the $y$-kinetic term of ${\cal H}$.

Since $\H(y)$ satisfies the periodic boundary condition, $\H(y)$ can be
decomposed as
\al{
{\H(y) = \sum_{n=-\infty}^{\infty} \frac{\vec{a}_n}{\sqrt{L}} e^{i 2 \pi n \frac{y}{L}},}
   \label{modH_profile}
}
where {} {$\vec{a}_n$ is a two-component $SU(2)_W$ constant vector}.
{Substituting Eq.~(\ref{modH_profile}) into $\E_1$, we obtain} the expression
\al{
\E_{1} = \sum_{n=-\infty}^{\infty}
   \left[ \left(\frac{2 \pi n + \theta}{L}\right)^2 - \left(\frac{\theta}{L}\right)^2 \right]
   {\ab{\vec{a}_n}^2 \geq 0}
}
and we can conclude that 
the minimum of $\E_{1}$ is given by $\E_{1}=0$
when the values of $\theta$ and {$\vec{a}_n$} satisfy one of the {conditions}
\al{
\text{(i)}\ &
{-\pi < \theta < \pi \text{\ and\ } {\vec{a}_n} = 0\,(n \not= 0):}
   &&\H = \frac{{\vec{a}_0}}{\sqrt{L}},
   \label{vevform_i} \\
\text{(ii)}\ &
\theta = \pi \text{\ and\ } {\vec{a}_n} = 0\,(n \not= 0,-1):
   &&{\H = \frac{\vec{a}_0}{\sqrt{L}} \ \ \text{or}\ \ \H = \frac{{\vec{a}_{-1}}}{\sqrt{L}}
   e^{-i 2\pi \frac{y}{L}},} \label{vevform_ii}
}
where {$\vec{a}_{0}$ in Eq.~(\ref{vevform_i}) and $\vec{a}_{0}$ or $\vec{a}_{-1}$ in Eq.~(\ref{vevform_ii})} are still undetermined.
The functional $\E_2$ 
takes the minimum value
{if the following 
condition is fulfilled:}
\al{
\ab{\H}^2 =
\begin{cases}
    {\frac{1}{\lambda}\left({M}^2 - \left(\frac{\theta}{L}\right)^2\right)}
   & \text{for\ }  {{M}^2} - \left(\frac{\theta}{L}\right)^2 > 0 \\
    0
   & \text{for\ }  {{M}^2} - \left(\frac{\theta}{L}\right)^2 \leq 0
\end{cases}.
}

Combining the above two results and using the $SU(2)_{W}$
global symmetry, we can show that the VEV $\langle H(y) \rangle$ is
given, without loss of generality, as
\begin{description}
\item[(I)] {${M}^2 - \left(\frac{\theta}{L}\right)^2 > 0$}
\al{
   \vev{H(y)} = 
   \begin{cases}
      \frac{v}{\sqrt{2}}\,
      e^{i {\frac{\theta}{L}y}}
      \begin{pmatrix} 0 \\ 1 \end{pmatrix}
      & \text{for\ } {-\pi < \theta < \pi}, \\
      \frac{v}{\sqrt{2}}\,
      e^{i {\frac{{\pi}}{L}y}}
      \begin{pmatrix} 0 \\ 1 \end{pmatrix}
      \text{\ \ or\ \ }
      \frac{v}{\sqrt{2}}\,
      e^{{-i \frac{{\pi}}{L}y} }
      \begin{pmatrix} 0 \\ 1 \end{pmatrix}
      & \text{for\ } {\theta = \pi}, \\
   \end{cases}
   \label{doubletVEVform1}
}
\item[(II)] {${M}^2 - \left(\frac{\theta}{L}\right)^2 \leq 0$}
\al{
\vev{H(y)} = \begin{pmatrix} 0 \\ 0 \end{pmatrix}{,} \label{doubletVEVform2}
}
\end{description}
where $v$ is given by
\al{
\left(\frac{v}{\sqrt{2}}\right)^2 :=
{\left|\vev{H(y)}\right|^2} = {\frac{1}{\lambda}\left({M}^2 - \left(\frac{\theta}{L}\right)^2\right)}.
\label{squared_VEV}
}
From now on, we will assume the case of (I) 
${M}^2 - \left(\frac{\theta}{L}\right)^2 > 0$.

Now we discuss some properties of the derived VEV in Eq.~(\ref{doubletVEVform1}). Differently from the SM, the VEV possesses {$y$-position dependence,} and its broken phase is realized only in the case of {${M}^2 - \left(\frac{\theta}{L}\right)^2 > 0$}{. But} like the SM,
{the squared VEV~(\ref{squared_VEV}) is still constant even though $\langle H(y) \rangle$ depends on $y$}.
This means that after {$v\sqrt{L}$} is set as $246\,\text{GeV}$, where the mass dimension of
$v$ is $3/2$, the same situation {as} the SM
occurs in the EWSB sector.
On the other hand, the {$y$ dependence} of the Higgs VEV
in {Eq.}~(\ref{doubletVEVform1}) {is} an important consequence for the Yukawa
sector.
Since the VEV of the Higgs doublet appears linearly in each Yukawa term,
the overlap integrals which lead to effective 4D Yukawa couplings 
will produce {a nontrivial $CP$} phase in the CKM matrix.

In terms of the VEV and physical Higgs modes $h^{(n)}(x)$,
$H$ {can be expanded} as
\al{
H(x,y) \rightarrow
	\sum_{n=-\infty}^{\infty}
	\begin{pmatrix}
	0 \\
	\frac{1}{\sqrt{2}} \left( v e^{i \frac{\theta}{L} y} \delta_{n,0} + h^{(n)}(x) \frac{1}{\sqrt{L}} e^{i \left( \frac{2 \pi n + \theta}{L} \right)y} \right)
	\end{pmatrix},
\label{doublet_KKexpansion}
}
which obeys the boundary condition (\ref{twistBC}).
The physical {masses} $\mu_{h^{(n)}}$ of the zero mode $(n=0)$ and the {Kaluza--Klein} (KK) modes $(n \not= 0)$ are easily calculated from Eq.~(\ref{doubletaction}) {as}
\al{
\mu_{h^{(n)}}^2 = 
	\begin{cases}
	2 \left( M^2 - \left(\frac{\theta}{L}\right)^2 \right) = \lambda v^2 & \text{for\ } n=0 \\
	{ 2{M}^2 + \frac{(\theta+2\pi n)^2}{2L^2} + \frac{(\theta-2\pi n)^2}{2L^2} -3 \left( \frac{\theta}{L} \right)^2  = {\lambda v^2} + \frac{(2\pi n)^2}{L^2}}  & \text{for\ } {n \geq 1}
	\end{cases},
\label{physicalHiggsmasssq}
}
with the hermiticity condition for a real field on $S^1$: $h^{(n)\dagger} = h^{(-n)}$.

We mention that the relation between {$\mu_{h^{(n)}}$ and $\lambda$} {for $n=0$} in Eq.~(\ref{physicalHiggsmasssq}) is totally the same as that of the {standard model}.
We also comment on the Higgs-quarks couplings in our model.
As shown in Eq.~(\ref{doublet_KKexpansion}), the profiles of the VEV and the Higgs physical zero mode
are the same as $e^{i \frac{\theta}{L} y}$ up to the coefficients.
This means that the strengths of the couplings are equivalent to those of the SM even though the mode function gets to be $y$-position dependent.
As a result, the decay branching ratios of the Higgs boson are the same as those of the {standard model}.\footnote{{Being different from the {universal extra dimension} case~\cite{Petriello:2002uu,Nishiwaki:2011vi,Nishiwaki:2011gk,Kakuda:2012px,Belanger:2012mc}, the ``low" KK mass less than a TeV scale is not allowed after considering the {level mixing} in the top sector~\cite{Fujimoto:2012wv}.
Then, the significant deviations do not occur
in the loop-induced single Higgs production via gluon fusion and Higgs decay processes to a pair of photons and gluons in our model.}
}

%%%%%%%%%%%%%%%%%%%%%%%%%%%%%%%%%%%%%%%%%%%%%
%%%%%%%%%%%%%%%%%%%%%%%%
\section{{Model} with point interactions on $S^1$}
%%%%%%%%%%%%%%%%%%%%%%%%
%%%%%%%%%%%%%%%%%%%%%%%%%%%%%%%%%%%%%%%%%%%%%

%%%%%%%%%%%%%%%%%%%%%%%%%%%%%%%%%
%\vspace{-4mm} %%%
%%%%%%%%%%%%%%%%%%%%%%%%%%%%%%%%%
%
\begin{figure}[t]
\centering
\includegraphics[width=0.99\columnwidth]{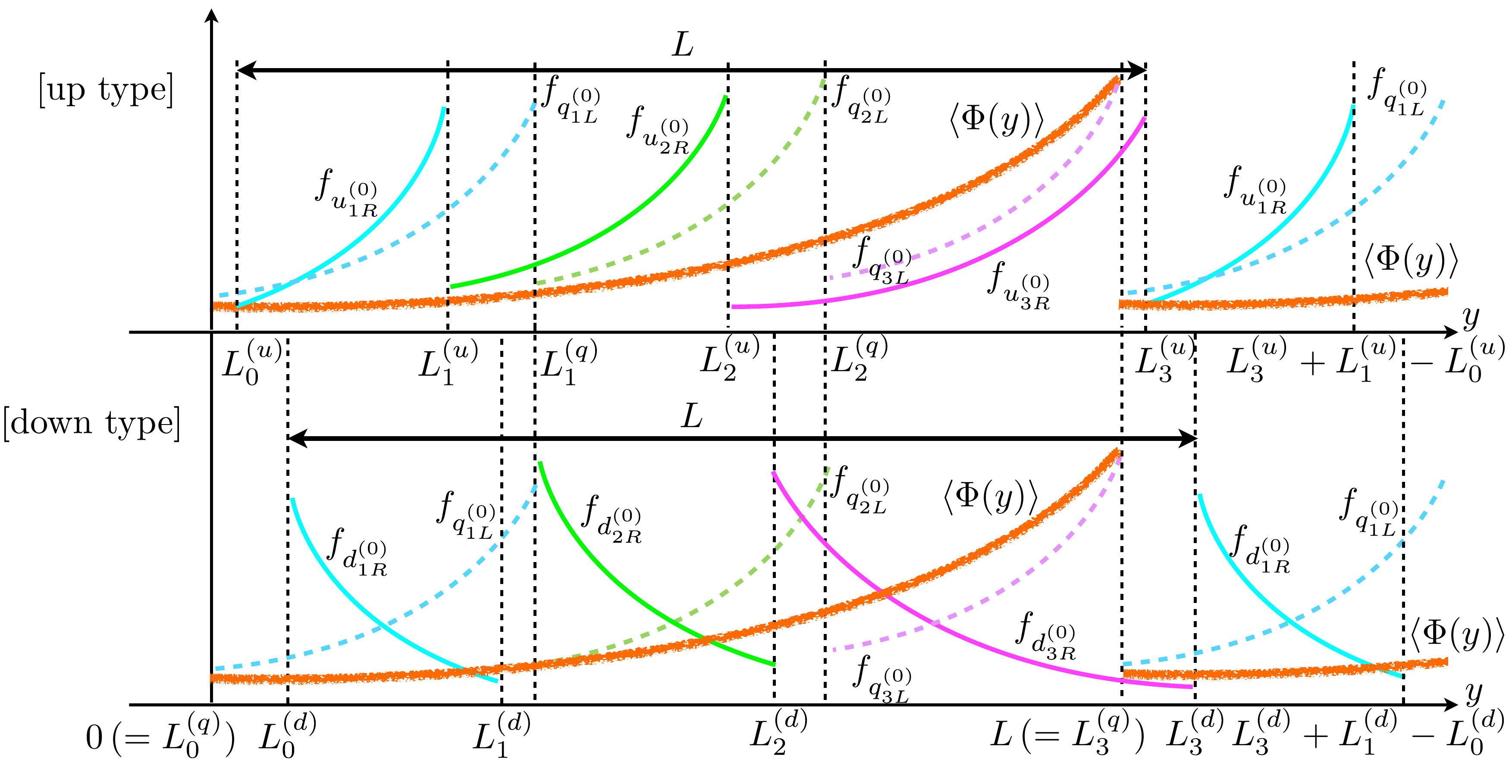}
\caption{
The {wave function} profiles of the quarks and the VEV of $\Phi(y)$ are
schematically depicted. 
Here we take $L_0^{(q)}=L_0^{(\Phi)}=0$.
Note that all the profiles
have the periodicity along $y$ with the same period $L$.
Differently from the model on an interval in Ref.~\cite{Fujimoto:2012wv}, we can find the $(1,3)$ elements of the mass matrices due to the periodicity along {the $y$ direction}.
}
\label{totalprofile_pdf}
\end{figure}

\begin{figure}[t]
\centering
\includegraphics[width=0.8\columnwidth]{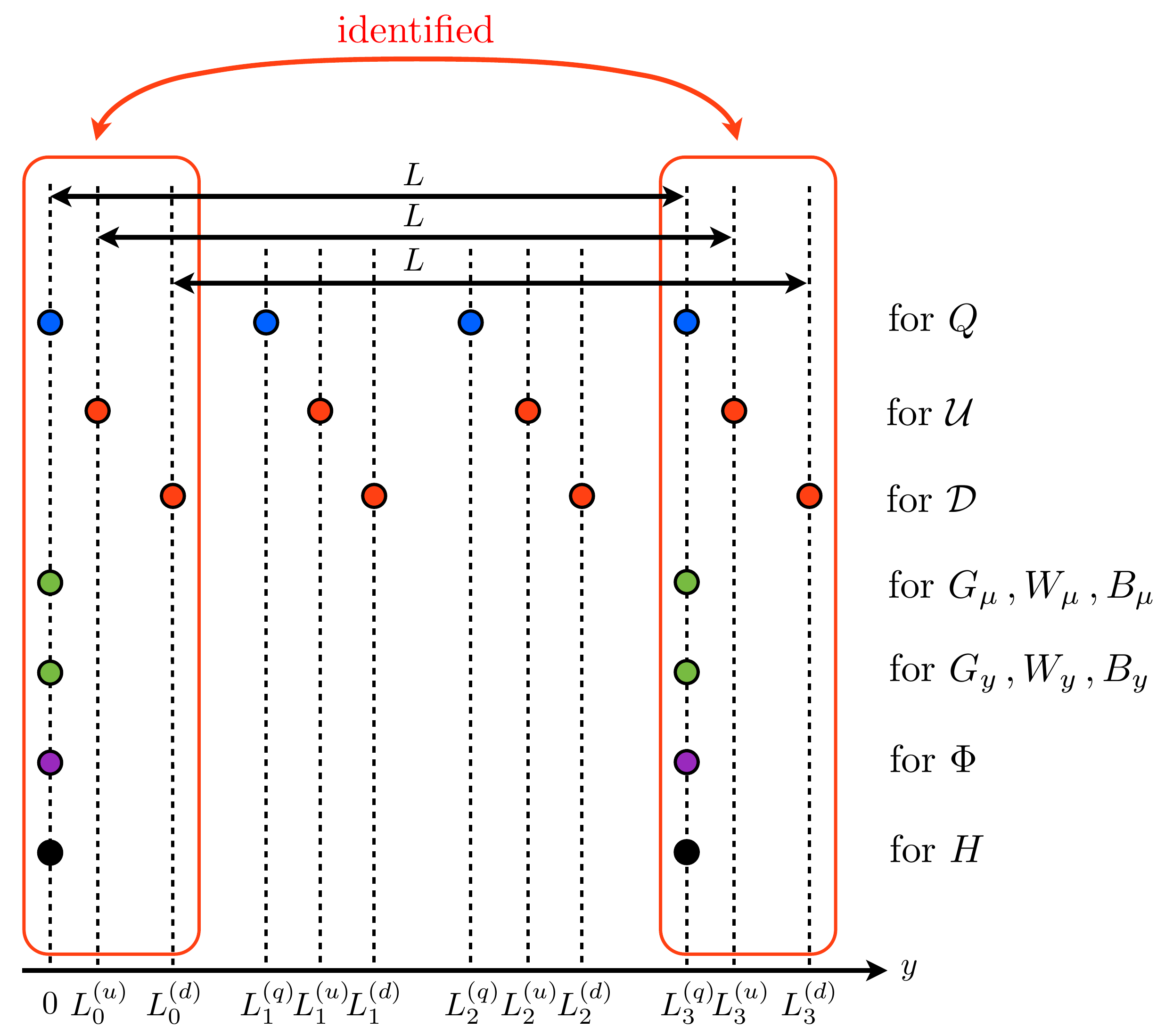}
\caption{
This is an overview of our system with point interactions. The red (blue) circular spots show the Dirichlet BC for left- (right-)handed part at the corresponding boundary points, respectively.
The green, purple,  and black circular spots represent the ordinary {periodic in Eq.~(\ref{gluon_BC})}, the Robin BCs in Eq.~(\ref{RobinBC}), and the twisted BCs in Eq.~(\ref{twistBC}), respectively.
{It is noted that we adopt the assumption in Eq.~(\ref{L_assumption}).}
}
\label{BCs_pdf}
\end{figure}

In the previous section, we introduced the twisted BC for the $SU(2)_W$ doublet $H$ and {generated} the EWSB by the {$y$-position-dependent} complex VEV in Eq.~(\ref{doubletVEVform1}) in the case of $M^2 - (\theta/L)^2 > 0$.
We expect that this VEV also works as the source of the {$CP$} phase of the CKM matrix, but here an important issue, which we should think {about} carefully, exists.

If all the profiles of the {{three-generation}} quarks are flat, an effective phase appears after integration over $y$ just as an overall factor, which can be removed by $U(1)$ rephasing and never works as a physical {$CP$} phase.
To circumvent this difficulty, profiles of the quarks are required to be localized.
On the other hand, field localization (in extra dimensions) is known as an effective way of explaining the quark mass hierarchy and pattern of flavor mixing.
In this section, we consider a model with point interactions as an illustrative example.
Point interaction can be considered as zero-thickness brane and we can arrange it anywhere in the bulk space of $S^1$.

At the location of a point interaction, we can consider {five-dimensional (5D)} gauge-invariant boundary conditions, {for which the variety is} rich compared with the case of $Z_2$ orbifolding.
After we introduce three point interactions for a 5D fermion, its zero-mode profile gets to be chiral, split and localized.
This situation is just what we want.\footnote{
{Another interesting idea for generating three-generation structure and field localization is introducing magnetic flux on {the} torus~\cite{Cremades:2004wa,Fujimoto:2013xha,Abe:2013bca}.}
}
We emphasize that flavor mixing is naturally realized as overlapping of localized quark profiles.
In the model, {an} additional gauge singlet scalar is required for generating the large mass hierarchy of the quarks.
Its (almost) exponential shape in {the} VEV is also generated by a suitable boundary condition at the corresponding point interactions.

This basic idea is found in Ref.~\cite{Fujimoto:2012wv}.
Nevertheless, there are two different points between the models 
in this {paper} and in {Ref.}~\cite{Fujimoto:2012wv}:

\begin{itemize}
\item
In the previous model \cite{Fujimoto:2012wv}, the Higgs VEV cannot 
possess a {nontrivial} complex phase, and a {$CP$} phase in the CKM matrix
has not been realized.
On the other hand, the VEV in our present model has a 
$y$-position-dependent complex phase, which will produce a {$CP$} phase
of the CKM matrix.
\item
In the previous model \cite{Fujimoto:2012wv}, the extra dimension has
been taken to be an interval, where the twisted BC in Eq. (\ref{twistBC})
cannot be realized.
In the present model, we set the extra dimension to be a circle
$S^{1}$, {for which the} geometry is compatible with the twisted BC (\ref{twistBC}).
\end{itemize}

In the following part, we briefly
explain how to construct our model.
The 5D action for fermions is
given by\footnote{
We adopt the representations of the gamma matrices {as}
$
\Gamma_\mu = \gamma_\mu, \ \Gamma_y = \Gamma^y = -i \gamma^5 = \gamma^0 \gamma^1 \gamma^2 \gamma^3
$
and  the Clifford algebra is defined as
$
\br{\Gamma_M, \Gamma_N} = -2 \eta_{MN}.
$
}
\al{
S &= \int d^4 x \int_{0}^{L} dy \Bigg\{ \Big[ \overline{Q} \paren{i \pal_M \Gamma^M + M_{Q} } Q + \overline{\mathcal{U}} \paren{i \pal_M \Gamma^M + M_{\mathcal{U}} } \mathcal{U} \notag \\
&\phantom{=\int d^4 x \int_{0}^{L_3} \Bigg\{ \ \, } + \overline{\mathcal{D}} \paren{i \pal_M \Gamma^M + M_{\mathcal{D}} } \mathcal{D}
\Big] \Bigg\},
\label{fermion_action} 
}
where we introduce 
an $SU(2)_W$ doublet ($Q$), an up-quark singlet ($\U$), 
and a down-type singlet ($\D$) 
with the corresponding bulk masses ($M_Q, M_\U, M_\D$).
{We note that our model contains only one generation 
for 5D quarks
but each 5D quark produces three generations of the 4D quarks,
as we will see below.}

We adopt the following BCs for $Q, \U, \D$
with an infinitesimal {positive} constant
$\varepsilon$ \cite{Fujimoto:2012wv}:
\al{
Q_R &= 0  \qquad \text{at} \quad y=L_0^{(q)} + \varepsilon,\, L_1^{(q)} \pm \varepsilon,\, L_2^{(q)} \pm \varepsilon,\, L_3^{(q)} - \varepsilon, \label{SMfermion_BC1} \\
\U_L &= 0  \qquad \text{at} \quad y=L_0^{(u)} + \varepsilon,\, L_1^{(u)} \pm \varepsilon,\, L_2^{(u)} \pm \varepsilon,\, L_3^{(u)} - \varepsilon, \label{SMfermion_BC2} \\
\D_L &= 0  \qquad \text{at} \quad y=L_0^{(d)} + \varepsilon,\, L_1^{(d)} \pm \varepsilon,\, L_2^{(d)} \pm \varepsilon,\, L_3^{(d)} - \varepsilon, \label{SMfermion_BC3}
}
where $\Psi_{R}$ and $\Psi_{L}$ denote the eigenstates of $\gamma^{5}$, {i.e.,}
$\Psi_{R} \equiv \frac{1+\gamma^{5}}{2}\Psi$ and
$\Psi_{L} \equiv \frac{1-\gamma^{5}}{2}\Psi$.
{Here $L^{(i)}_j$ for $i=q, u, d$ and $j=0,1,2,3$ means the positions of point interactions for the 5D fermions.
See Figs.~\ref{totalprofile_pdf} and \ref{BCs_pdf} for details.}
A crucial consequence of the above BCs is that there appear 
{threefold} degenerated left- {(right-)handed} zero modes in the 
mode expansions of $Q$ ($\U,\D$) and that they form the three
generations of the quarks.
The details have been given in Ref.~\cite{Fujimoto:2012wv}.
We will not repeat the discussions here.

The fields $Q, \mathcal{U}, \mathcal{D}$  with the BCs in Eqs~(\ref{SMfermion_BC1})--(\ref{SMfermion_BC3}) are {KK decomposed} as follows:
\al{
Q(x,y) = \begin{pmatrix} U(x,y) \\ D(x,y) \end{pmatrix} &=
\begin{pmatrix}
\sum_{i=1}^{3} u^{(0)}_{iL}(x) f_{q^{(0)}_{iL}}(y) \\
\sum_{i=1}^{3} d^{(0)}_{iL}(x) f_{q^{(0)}_{iL}}(y)
\end{pmatrix}
+ (\text{KK modes}), \\
\mathcal{U}(x,y) &=
\sum_{i=1}^{3} u^{(0)}_{iR}(x) f_{u^{(0)}_{iR}}(y) + (\text{KK modes}), \\
\mathcal{D}(x,y) &=
\sum_{i=1}^{3} d^{(0)}_{iR}(x) f_{d^{(0)}_{iR}}(y) + (\text{KK modes}).
}
Here the {zero-mode} functions are obtained in the following forms:
\al{
f_{q^{(0)}_{iL}}(y) &= 
		\mathcal{N}_{i}^{(q)} e^{M_{Q} (y-L_{i-1}^{(q)})} \Big[ \theta(y-L_{i-1}^{(q)}) \theta(L_{i}^{(q)}-y) \Big]
		&& \text{in\ }[L_0^{(q)},L_3^{(q)}], \label{doubletwavefunction} \\
f_{u^{(0)}_{iR}}(y) &= 
		\mathcal{N}_{i}^{(u)} e^{-M_{\mathcal{U}} (y-L_{i-1}^{(u)})} \Big[ \theta(y-L_{i-1}^{(u)}) \theta(L_{i}^{(u)}-y) \Big]
		&& \text{in\ }[L_0^{(u)},L_3^{(u)}], \label{upsingletwavefunction} \\
f_{d^{(0)}_{iR}}(y) &= 
		\mathcal{N}_{i}^{(d)} e^{-M_{\mathcal{D}} (y-L_{i-1}^{(d)})} \Big[ \theta(y-L_{i-1}^{(d)}) \theta(L_{i}^{(d)}-y) \Big]
		&& \text{in\ }[L_0^{(d)},L_3^{(d)}], \label{downsingletwavefunction}
}
where
\al{
\Delta L^{(l)}_{i} &= L^{(l)}_{i} - L^{(l)}_{i-1} \qquad
(\text{for } i=1,2,3;\ l=q,u,d),
}
\vspace{-8mm}
\al{
\mathcal{N}_{i}^{(q)} = \sqrt{\frac{2M_{Q}}{e^{2M_{Q} \Delta L_{i}^{(q)} }-1}}, \quad
\mathcal{N}_{i}^{(u)} = \sqrt{\frac{2M_{\mathcal{U}}}{1-e^{-2M_{\mathcal{U}} \Delta L_{i}^{(u)} }}}, \quad
\mathcal{N}_{i}^{(d)} = \sqrt{\frac{2M_{\mathcal{D}}}{1-e^{-2M_{\mathcal{D}} \Delta L_{i}^{(d)} }}}.
\label{wavefunctionconventions}
}
$\mathcal{N}_{i}^{(q)}, \mathcal{N}_{i}^{(u)}, \mathcal{N}_{i}^{(d)}$ are the {wave function} normalization factors for $f_{q^{(0)}_{iL}}, f_{u^{(0)}_{iL}}, f_{d^{(0)}_{iL}}$, respectively.

Since the length of the total system is universal, 
$L^{(l)}_{3} - L^{(l)}_{0}\ (l=q,u,d)$ should be equal to the
circumference of $S^{1}$, i.e.
\al{
L := L^{(q)}_{3} - L^{(q)}_{0} = L^{(u)}_{3} - L^{(u)}_{0} = L^{(d)}_{3} - L^{(d)}_{0}.
\label{totallength}
}
Note that all the mode functions in Eqs.~(\ref{doubletwavefunction})--(\ref{downsingletwavefunction}) (and a form of a singlet VEV in Eq.~(\ref{exactHiggsVEVform})) are periodic with the common period $L$, whereas we do not indicate that thing explicitly in Eqs.~(\ref{doubletwavefunction})--(\ref{downsingletwavefunction}).

In this model, the large mass hierarchy is naturally explained with {the Yukawa sector}
\al{
S_{\Y} = \int d^4 x \int_{0}^{{L}} dy \bigg\{ \Phi \Big[
		- \Y^{(u)} \overline{Q} (i \sigma_2 H^{\ast}) \U - \Y^{(d)} \overline{Q} H \D  \Big]
		+ \text{h.c.}
		\bigg\},
\label{higherdimensionalYukawa}
}
where $\Y^{(u)}/\Y^{(d)}$ is the Yukawa coupling for {up-/down-type} quark;
$H$ and $\Phi$ are an $SU(2)_W$ scalar doublet and a singlet.
It should be noted that 
although the Yukawa couplings $\Y^{(u)}$ and $\Y^{(d)}$ can be complex, 
they cannot be an origin of the {$CP$} phase of the CKM matrix
because our model contains only a single quark generation {so that}
the number of the 5D Yukawa couplings is not enough
to produce a {$CP$} phase in the CKM matrix.
An outline of our system is {depicted} in Fig.~\ref{totalprofile_pdf}.
Note that the five terms of $\overline{Q} (i \sigma_2 H^{\ast}) \U, \overline{Q} H \D, \Phi \overline{Q} Q, \Phi \overline{\U} \U, \Phi \overline{\D} \D$ with the Pauli matrix $\sigma_2$ are
excluded by introducing a discrete symmetry 
$H \rightarrow -H, \Phi \rightarrow -\Phi$.
$\Phi$ is a gauge singlet and there is no problem with gauge universality violation.\footnote{
If there exists the doublet-singlet mixing term $- C H^{\dagger} H \Phi^{\dagger} \Phi$ with a coefficient $C$, which cannot be prohibited by the discrete symmetry $H \to -H, \Phi \to -\Phi$ 
in our theory, gauge universality violation should be revisited.
A bound from the universality in {$Z$ boson gauge couplings} was already calculated as $CL \lesssim 0.003$ {(when a KK scale is around a few TeV)} in a model on an interval \cite{Fujimoto:2012wv}.
In this {paper}, we simply ignore this term.
}

The 5D action and the BCs for $\Phi$ are
assumed to be of the form \cite{Fujimoto:2012wv,Fujimoto:2011kf}
\al{
S_{\Phi} = \int d^4 x \int_0^{{L}} dy \bigg\{ {\Phi^{\dagger} \paren{\pal_M \pal^M - {M_{\Phi}}^2} \Phi
		- \frac{\lambda_{\Phi}}{2} \paren{\Phi^\dagger \Phi}^2} \bigg\}, 
\label{Phi_action}
}
\vspace{-8mm}
\al{
\Phi + L_+ \pal_y \Phi &= 0 \qquad \text{at} \quad y=L_{0}^{(\Phi)} + \varepsilon, \notag \\
\Phi - L_- \pal_y \Phi &= 0 \qquad \text{at} \quad y=L_3^{(\Phi)} - \varepsilon,
\label{RobinBC}
}
where {$M_\Phi$ ($\lambda_\Phi$)} is the bulk mass (quartic coupling) of the scalar singlet $\Phi$ and $L_{\pm}$ can take values in the range of $-\infty \leq L_{\pm} \leq \infty$ and $L_{0}^{(\Phi)}$ and $L_{3}^{(\Phi)}$ indicate the locations of the two ``end points" of the singlet.

The VEV of $\Phi$ with the BCs, named Robin BCs, in Eq.~(\ref{RobinBC}) is expressed in terms of Jacobi's elliptic functions in general and its phase structure {has been discussed} in Ref~\cite{Fujimoto:2011kf}.
We adopt a specific form {in the region $[L_{0}^{(\Phi)} + \varepsilon, L_{3}^{(\Phi)} - \varepsilon]$} \cite{Fujimoto:2012wv}{,}
\al{
\vev{\Phi(y)} = \sqbr{ {\frac{M_\Phi}{\sqrt{\lambda_\Phi}}} \br{\sqrt{1 + X}-1}^{1/2} }
\times  \frac{1}{ \text{cn}  \paren{ {M_\Phi} \br{1 + X}^{1/4} (y-y_0), \sqrt{\frac{1}{2}\paren{1 + \frac{1}{\sqrt{1+X}}}}       }  },
\label{exactHiggsVEVform}
}
with
\al{
X:= {\frac{4\lambda_\Phi |Q|}{M^4_\Phi}}.%}.
\label{parameter_k}
}
Here $y_0$ and $Q$ are parameters which appear after integration on $y$ and we focus on the choice of $Q<0$.
We note that the values of $y_0$ and $Q$ are automatically determined after choosing those of $L_{\pm}$.
As shown in Ref.~\cite{Fujimoto:2012wv}, 
we get the form of $\vev{\Phi(y)}$ to be an (almost) exponential function
of $y$ by
choosing suitable parameter configurations.
{Although there is a discontinuity in the {wave function} profile of $\langle \Phi \rangle$ between $y=L_{0}^{(\Phi)}+\varepsilon$ and $y=L_{3}^{(\Phi)}-\varepsilon$ in Eqs.~(\ref{RobinBC}),
this type of {BC} is derived from the variational principle on $S^1$
and leads to no inconsistency~\cite{Fujimoto:2011kf}.}

The BCs for the 5D $SU(3)_C, SU(2)_W, U(1)_Y$ gauge bosons $G_M, W_M, B_M$ are selected as
\al{
{G_{M}|_{y=0} = G_{M}|_{y=L},\quad
\pal_y G_{M}|_{y=0} = \pal_y G_{M}|_{y=L},}
\label{gluon_BC}
}
where we only show the $G_M$'s case.
In this configuration, we obtain the {standard model} gauge bosons in zero modes.
Based on the discussion in {Sec.}~\ref{section_2}, we conclude that the W and Z bosons become massive and their masses are suitably created through ``our" Higgs mechanism as $m_W \simeq 81\,\text{GeV}, m_Z \simeq 90\,\text{GeV}$.
The overview of the BCs is summarized in Fig.~\ref{BCs_pdf}.
We mention that, on $S^1$ geometry, $G^{(0)}_y$, $W^{(0)}_y$, and $B^{(0)}_y$
would exist as massless 4D scalars at the tree level, 
but they will become massive via quantum corrections and are
expected to be uplifted to near KK states. We will discuss those modes
in another paper.
We should note that in our model on $S^1$ with point interactions, the 5D gauge symmetries are intact under the BCs summarized in Fig.~\ref{BCs_pdf}.\footnote{
{In Refs.~\cite{Lim:2005rc,Lim:2007fy,Lim:2008hi}, the 5D gauge invariance has been discussed
from a quantum mechanical supersymmetry point of view.}
}
Hence the unitarity in the scattering processes of massive particles are ensured in our model.\footnote{
Some related works are found in Refs.~\cite{SekharChivukula:2001hz,Abe:2003vg,Chivukula:2003kq,Csaki:2003dt,Ohl:2003dp,Abe:2004wv,Sakai:2006qi,Nishiwaki:2010te}.}

%%%%%%%%%%%%%%%%%%%%%%%%%%%%%%%%%%%%%%%%%%%%%
%%%%%%%%%%%%%%%%%%%%%%%%
\section{{$CP$} phase in the CKM matrix}
%%%%%%%%%%%%%%%%%%%%%%%%
%%%%%%%%%%%%%%%%%%%%%%%%%%%%%%%%%%%%%%%%%%%%%

In this section, we verify that our mechanism can actually produce a
{nontrivial $CP$} phase
in the CKM matrix. We further would like to find a set of parameter configurations
in which the quark mass hierarchy and the structure of the CKM matrix
are derived naturally.
In the following analysis, we
rescale all the
dimensional valuables by the $S^1$ circumference $L$ 
to make them dimensionless and the
rescaled valuables are indicated with the tilde $\tilde{{}}$\,.

We set the parameters concerning the scalar singlet $\Phi$ as
\al{
{\tilde{M}_\Phi} = 8.67, \quad \tilde{y}_0 = - 0.1, \quad {\tilde{\lambda}_\Phi} = 0.001, \quad |\tilde{Q}| = 0.001,
\label{asetofHiggsparameters}
}
where the VEV profile becomes an (almost) exponential
function of $y$, 
which is suitable for generating the large mass hierarchy.\footnote{
The smallness of $Q$ is not an unnatural thing {because} they are resultant values derived from the two input parameters $L_{\pm}$, {for which the} dimensionless values are within $\mathcal{O}(10)$ as in Eq.~(\ref{Lplusminusvalues}).
{We note that $\lambda_\Phi$ always appears in the form of the singlet VEV in Eq.~(\ref{exactHiggsVEVform}) as the combination $|Q|\lambda_\Phi$. %{.} and
$\lambda_\Phi$ in itself only affects the overall normalization.
Therefore some room might remain for {a} more ``natural" choice of $\lambda_\Phi$.}
}
In this case, the values of $L_{\pm}$ %describing the singlet's BC 
in Eq.~(\ref{RobinBC}) correspond to
\al{
\frac{1}{\tilde{L}_+} = - 6.07, \quad \frac{1}{\tilde{L}_-} = 8.69,
\label{Lplusminusvalues}
}
where the broken phase is realized %in the singlet~
\cite{Fujimoto:2012wv}.

\begin{table}[t]
   \centering
  \begin{tabular}{|c|cc||c|cc|} \hline
    $M_{ij}^{(u)}$ & $a$ & $b$ & $M_{ij}^{(d)}$ & $a$ & $b$ \\ \hline \hline
    $M_{11}^{(u)}$ & $L_0^{(u)}$ & $L_1^{(u)}$ & $M_{11}^{(d)}$ & $L_0^{(d)}$ & $L_1^{(d)}$ \\
    $M_{22}^{(u)}$ & $L_1^{(q)}$ & $L_2^{(u)}$ & $M_{22}^{(d)}$ & $L_1^{(q)}$ & $L_2^{(d)}$ \\
    $M_{33}^{(u)}$ & $L_2^{(q)}$ & $L$ & $M_{33}^{(d)}$ & $L_2^{(q)}$ & $L$ \\
    $M_{12}^{(u)}$ & $L_1^{(u)}$ & $L_1^{(q)}$ & $M_{12}^{(d)}$ & $L_1^{(d)}$ & $L_1^{(q)}$ \\
    $M_{23}^{(u)}$ & $L_2^{(u)}$ & $L_2^{(q)}$ & $M_{23}^{(d)}$ & $L_2^{(d)}$ & $L_2^{(q)}$ \\
    $M_{31}^{(u)}$ & $L$ & $L+L_0^{(u)}$ & $M_{31}^{(d)}$ & $L$ & $L+L_0^{(d)}$ \\
    \hline
  \end{tabular}
\caption{The summary table for the overlap integrals in Eq.~(\ref{general_overlapintegral}).}
\label{integration_parameters}
\end{table}

As in the previous analysis~\cite{Fujimoto:2012wv}, the signs of the fermion bulk masses are assigned as
$M_Q > 0, M_\U < 0, M_\D > 0$ to make much larger overlapping in {the up-quark} sector than in down ones for top mass.
Here we assume the positions of the two {end points} of both the quark doublet and the scalar singlet are the same
\al{
L_0^{(q)} = L_0^{(\Phi)} = 0,\quad
L_3^{(q)} = L_3^{(\Phi)} = L,
\label{L_assumption}
}
where we set $L_0^{(q)}$ and $L_0^{(\Phi)}$ as zero.
In addition, we also assume that the orders of the positions of point interactions are settled as
\al{
0 < L_0^{(u)} < L_1^{(u)} < L_1^{(q)} < L_2^{(u)} < L_2^{(q)} < L < L_3^{(u)}, \notag \\
0 < L_0^{(d)} < L_1^{(d)} < L_1^{(q)} < L_2^{(d)} < L_2^{(q)} < L < L_3^{(d)}.
\label{positionorder1}
}

Here our {up-quark} mass matrix ${\mathcal{M}}^{(u)}$ and that of down ones ${\mathcal{M}}^{(d)}$ take the forms
\al{
\mathcal{M}^{(u)} =
\begin{bmatrix}
M^{(u)}_{11} & M^{(u)}_{12} & M^{(u)}_{13} \\
0 & M^{(u)}_{22} & M^{(u)}_{21} \\
0 & 0 & M^{(u)}_{33}  
\end{bmatrix}, \quad
\mathcal{M}^{(d)} =
\begin{bmatrix}
M^{(d)}_{11} & M^{(d)}_{12} & M^{(d)}_{13} \\
0 & M^{(d)}_{22} & M^{(d)}_{21} \\
0 & 0 & M^{(d)}_{33}  
\end{bmatrix},
\label{massmatrix1}
}
where the row (column) index of the mass matrices {shows the generations of the left- (right-)handed fermions}, respectively.
Differently from the model on an interval in Ref.~\cite{Fujimoto:2012wv}, {the $(1,3)$ elements of the mass matrices are allowed geometrically} due to the periodicity along {the $y$ direction}.
The general form of the nonzero matrix elements of ${\mathcal{M}}^{(u)}$ and ${\mathcal{M}}^{(d)}$
can be expressed {as}
\al{
M^{(\kappa)}_{ij} = {\Y}^{(\kappa)} \int_{a}^{b} \!\! dy 
		f_{q^{(0)}_{iL}}(y) f_{\kappa^{(0)}_{jR}}(y) \langle \Phi(y) \rangle
		{\langle H(y) \rangle},
\label{general_overlapintegral}
}
where {$\kappa$ indicates {the} up/down type of quark and} the concrete information is stored in Table~\ref{integration_parameters}.

%%%%%%%%%%%%%%%%%%%%%%%%%%
\subsection{{Quark masses and mixing parameters}
\label{results}}
%%%%%%%%%%%%%%%%%%%%%%%%%%

The parameters which we use for calculation are
\beq
\begin{array}{llll}
\tilde{L}_{0}^{(q)} = 0, & \tilde{L}_{1}^{(q)} = {0.30}, & \tilde{L}_{2}^{(q)} = {0.660}, &
		\tilde{L}_{3}^{(q)} = 1,  \\
\tilde{L}_{0}^{(u)} =  {0.024}, & \tilde{L}_{1}^{(u)} =  {0.026}, & \tilde{L}_{2}^{(u)} =  {0.52}, &
		\tilde{L}_{3}^{(u)} =  {1.024},  \\
\tilde{L}_{0}^{(d)} =  {0.07}, &
		\tilde{L}_{1}^{(d)} =  {0.18}, & \tilde{L}_{2}^{(d)} =  {0.646}, &
		\tilde{L}_{3}^{(d)} = {1.07}, \\
\tilde{M}_Q =  {6}, & \tilde{M}_\U =  {-6}, & \tilde{M}_\D =  {5}, &
		\theta =  {3},
\end{array}
\label{a_setofsolution}
\eeq
where the twist angle $\theta$ is a {dimensionless} value and should be within the range
$-\pi < \theta \leq \pi$.
We note that $\tilde{L}_3^{(q)}$, $\tilde{L}_3^{(u)}${,} and $\tilde{L}_3^{(d)}$ are considered to be not independent degree of freedom, {for which the} values are automatically determined after we choose the other positions of the point interactions.
In {Appendix}~\ref{appendix}, we will comment on the orders of significant digits of the input parameters in Eq.~(\ref{a_setofsolution}).
We should {note} that in our system, the EWSB is only realized on the condition of {${M}^2 - \left(\frac{\theta}{L}\right)^2 > 0$} as in Eqs.~(\ref{doubletVEVform1}).
Recently, the ATLAS and CMS experiments have announced that the physical Higgs mass is around $126\,\text{GeV}$ {over} $5\sigma$ confidence level~\cite{:2012gk,:2012gu}.
{$\tilde{\lambda}$} is $0.262$ irrespective of the value of $L$, {while {$\tilde{M}$} is slightly dependent on the value of $L$ as $3.01303$ $(3.00052)$ in the case of $M_{\text{KK}}=2\,\text{TeV}$ ($M_{\text{KK}}=10\,\text{TeV}$), where $M_{\text{KK}}$ is a typical scale of the KK mode and defined as $2\pi/L$}. 
Here some tuning is required to obtain the suitable values realizing the EWSB.

After the diagonalization of the two mass matrices, the quark masses are evaluated as
\begin{equation}
\begin{array}{lll}
m_{\text{up}} =  {2.5}\,\text{MeV}, &
	m_{\text{charm}} =  {1.339}\,\text{GeV}, &
	m_{\text{top}} =  {173.3}\,\text{GeV},  \\
m_{\text{down}} =  {4.8}\,\text{MeV}, &
	m_{\text{strange}} =  {104}\,\text{MeV}, &
	m_{\text{bottom}} =  {4.183}\,\text{GeV},  \\
\vspace{-4mm} \\
\dis \frac{m_{\text{up}}}{m_{\text{up}}|_{\text{exp.}}} =  {1.07}, &
	\dis \frac{m_{\text{charm}}}{m_{\text{charm}}|_{\text{exp.}}} =  {1.05}, &
	\dis \frac{m_{\text{top}}}{m_{\text{top}}|_{\text{exp.}}} = 1.00, \\
\vspace{-4mm} \\
\dis \frac{m_{\text{down}}}{m_{\text{down}}|_{\text{exp.}}} =  {0.993}, &
	\dis \frac{m_{\text{strange}}}{m_{\text{strange}}|_{\text{exp.}}} =  {1.10}, &
	\dis \frac{m_{\text{bottom}}}{m_{\text{bottom}}|_{\text{exp.}}} = 1.00,
\end{array}
\label{obtainedquarkmass}
\end{equation}
and the absolute values of the {CKM matrix elements} are given as\footnote{
{The values of $\tilde{\Y}^{(u)}$ and $\tilde{\Y}^{(d)}$ are also chosen as $\tilde{\Y}^{(u)}=-0.0532+0.0156\,i$ and $\tilde{\Y}^{(d)} = -0.00335-0.00146\,i$
by setting the initial conditions $M_{33}^{(u)} = m_t$ and  $M_{33}^{(d)} = m_b$ with the top mass $m_t$ and the bottom mass $m_b$.}
}
\al{
|V_{\text{CKM}}|=
\begin{bmatrix}
		 {0.971} &  {0.238} &   {0.00377} \\
		 {0.237} &  {0.971} &   {0.0403} \\
		 {0.00887} &  {0.0395} &  0.999
\end{bmatrix},\quad
\left| \frac{V_{\text{CKM}}}{V_{\text{CKM}}|_{\text{exp.}}} \right|=
\begin{bmatrix}
		0.997 & 1.06 &   {1.07} \\
		1.06 & 0.998 &   {0.978} \\
		 {1.02} &  {0.978} & 1.00
\end{bmatrix}.
\label{obtainedCKMmatrix}
}

%%%%%%%%%%%%%%%%%%%%%%%%%%
\subsection{{$CP$} phase
\label{results_CP}}
%%%%%%%%%%%%%%%%%%%%%%%%%%

The Jarlskog parameter $J$ containing information about the {$CP$} phase is defined by
\al{
\text{Im} \left[ (V_{\text{CKM}})_{ij} (V_{\text{CKM}})_{kl} (V_{\text{CKM}}^\ast)_{il} (V_{\text{CKM}}^\ast)_{kj} \right] = J \sum_{m,n=1}^{3} \epsilon_{ikm} \epsilon_{jln}
}
with the completely antisymmetric tensor $\epsilon$, {and is} invariant under the $U(1)$ unphysical {rephasing} operations of six types of quarks~\cite{Jarlskog:1985ht,Jarlskog:1985cw}.
This value is easily estimated as
\al{
J =  {3.23} \times 10^{-5},\quad \frac{J}{J|_{\text{exp.}}} =  {1.09},
}
where we also provide the differences from the latest experimental values in Ref.~\cite{Beringer:1900zz}.
All the deviations from the latest experimental values are within about {$10\%$,} and we can conclude that the situation of the SM is suitably generated.
In {Appendix}~\ref{appendix}, we discuss distribution patterns of quark mass-matrix elements and required orders in tuning the input parameters with the results for realizing the accuracy.

%%%%%%%%%%%%%%%%%%%%%%%%%%%%%%%%%%%%%%%%%%%%%
%%%%%%%%%%%%%%%%%%%%%%%%
\section{Summary and discussion}
%%%%%%%%%%%%%%%%%%%%%%%%
%%%%%%%%%%%%%%%%%%%%%%%%%%%%%%%%%%%%%%%%%%%%%

In this {paper}, we have proposed a new mechanism for generating {a $CP$ phase via a} Higgs VEV originating from {the} geometry of an extra dimension. A twisted BC for the Higgs doublet has been found to lead to an {extra-dimension coordinate-dependent} phase in the Higgs VEV, which contains a {nontrivial $CP$} phase degree of freedom. This mechanism is useful for generating a {$CP$} phase to a single generation {extra-dimensional} field theory incorporating with a generation production mechanism.
The electroweak symmetry breaking is also generated dynamically due to the twisted boundary condition with the suitable W- and Z-boson masses.

As an illustrative example, we applied our mechanism to a five-dimensional gauge theory on a circle with point interactions~\cite{Fujimoto:2012wv}. Point interactions, which are additional boundary points with respect to the extra dimension, are responsible for producing the three generations while the model consists of a {single-generation fermion} and make the quark profiles be localized. Since each element of the mass matrices picks up a different phase through the overlap integrals, there is some possibility of realizing a {nontrivial $CP$} phase.
After numerical calculations, we found that a {nontrivial $CP$} phase appears with good precision, maintaining the property of the original model in which the generations, the quark mass hierarchy and the CKM matrix appear from the geometry of the extra dimension. Certainly, our new mechanism for generating {a $CP$} phase via Higgs VEV works.

A key point of our mechanism is that we can generate both the EWSB and a {$CP$} phase simultaneously as a complex Higgs VEV.
To make our {$CP$-violation} mechanism work correctly, quark profiles should be split and localized.
In this situation, flavor mixing and mass hierarchy of the quarks are also naturally activated.
We would like to emphasize that in the model adopting our mechanism, all the concepts of quark flavor in the SM, namely {EWSB, the} number of generations, flavor mixing, mass hierarchy{, and $CP$} violation, are interlinked closely.

One of the most important remaining tasks is to construct a model which brings both the quarks and the leptons into perspective. Using our mechanism, not only the quark sector but also the lepton sector can acquire a {nontrivial $CP$} phase. Since the origin of the {$CP$} phase is common, we can predict the value of the {$CP$} phase of the lepton sector after fitting the value of the {$CP$} phase in the quark sector. The result will be reported elsewhere. Accommodation of our mechanism to another single generation model is also {an} important task.

Another crucial {topic} is the stability of the system.
Our system {is} possibly threatened with instability. 
Some {mechanisms} will be required to stabilize the moduli representing the positions of point interactions (branes).\footnote{
Moduli stabilization via Casimir energy in the system where a scalar takes the Robin BCs (but no point interaction in the bulk) has been studied in Refs.\cite{de_Albuquerque:2003uf,Bajnok:2005dx,Pawellek:2008st}.}
In a {multiply connected} space of $S^{1}$, there is another
origin of gauge symmetry breaking{, i.e.,} the Hosotani mechanism~\cite{Hosotani:1983xw,Hosotani:1988bm}.
Since further gauge symmetry breaking causes a problem
in the model, we need to insure that the Hosotani mechanism
does not occur. To this end, we might introduce additional
5D matter to prevent zero modes of {$y$ components} of gauge
fields from acquiring {nonvanishing} VEVs. We will leave
those issues in future work.

%%%%%%%%%%%%%%%%%%%%%%%%%%%%%%%%%%%%%%%%%%%%%%%%%%%%%%%
%%%%%%%%%%%%%%%%%%%%%%%%%%%%%%%%%%%%%%%%%%%%%%%%%%%%%%%
%%%%%%%%%%%%%%%%%%%%%%%%%%%%%%%%%%%%%%%%%%%%%%%%%%%%%%%
%\vskip 10pt
%\noindent
%{\large \bf Acknowledgments}
%\vskip 7pt
%\noindent
\section*{{Acknowledgments}}

The authors would like to thank T.Kugo for valuable discussions.
K.N. is partially supported by funding available from the Department of 
Atomic Energy, Government of India for the Regional Centre for Accelerator-based
Particle Physics (RECAPP), Harish-Chandra Research Institute.
This work is supported in part by a Grant-in-Aid for Scientific Research {[Grants No.\,22540281 and No.\,20540274 (M.S.)]} from the Japanese Ministry of Education, Science, Sports and Culture.

\appendix

\section*{Appendix}

%%%%%%%%%%%%%%%%%%%%%%%%%%%%%%%%%%%%%
\section{{Considering input-parameter dependence}
\label{appendix}}
%%%%%%%%%%%%%%%%%%%%%%%%%%%%%%%%%%%%%

\begin{figure}[t]
\centering
\includegraphics[width=0.40\columnwidth]{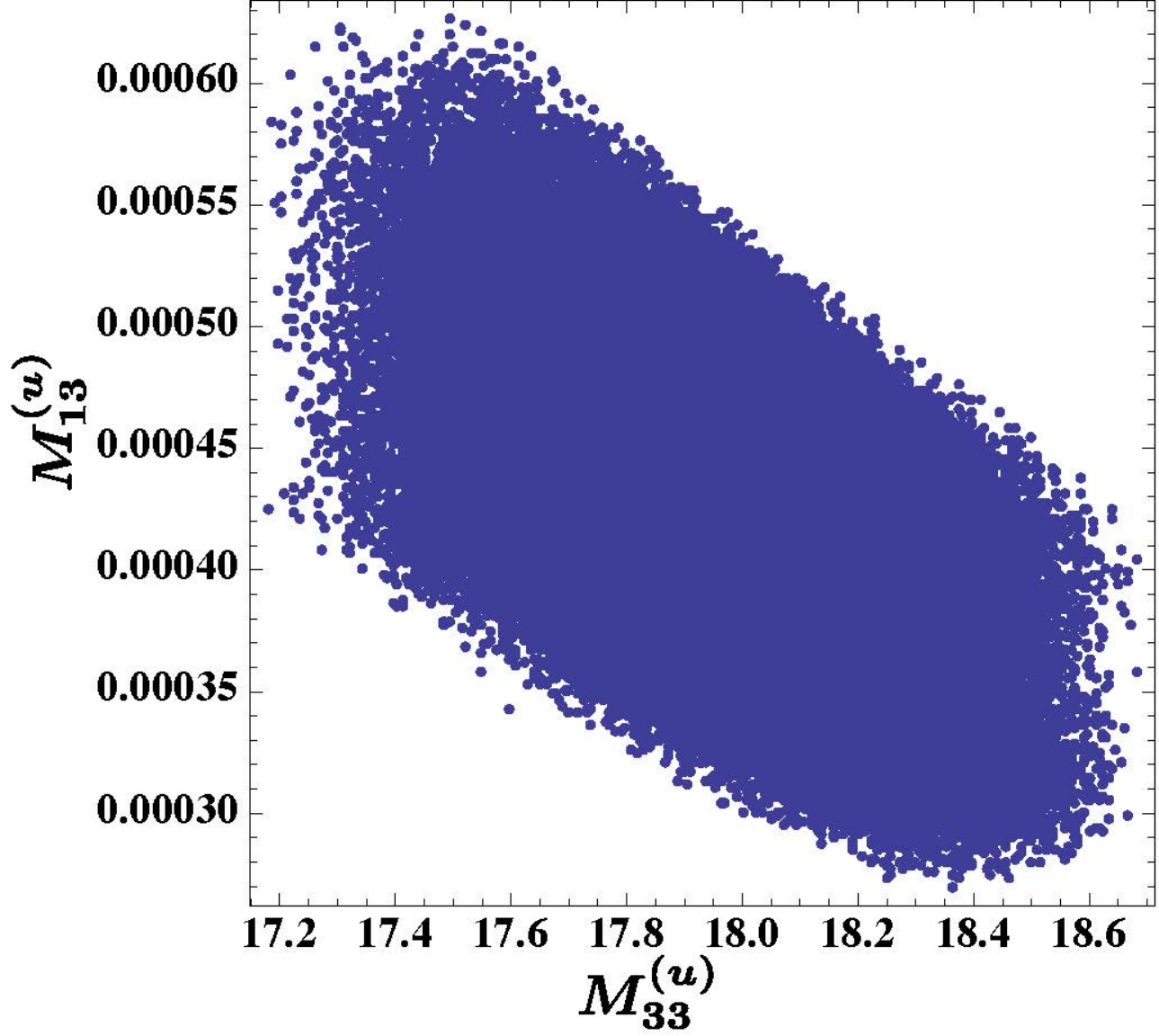}
\includegraphics[width=0.38\columnwidth]{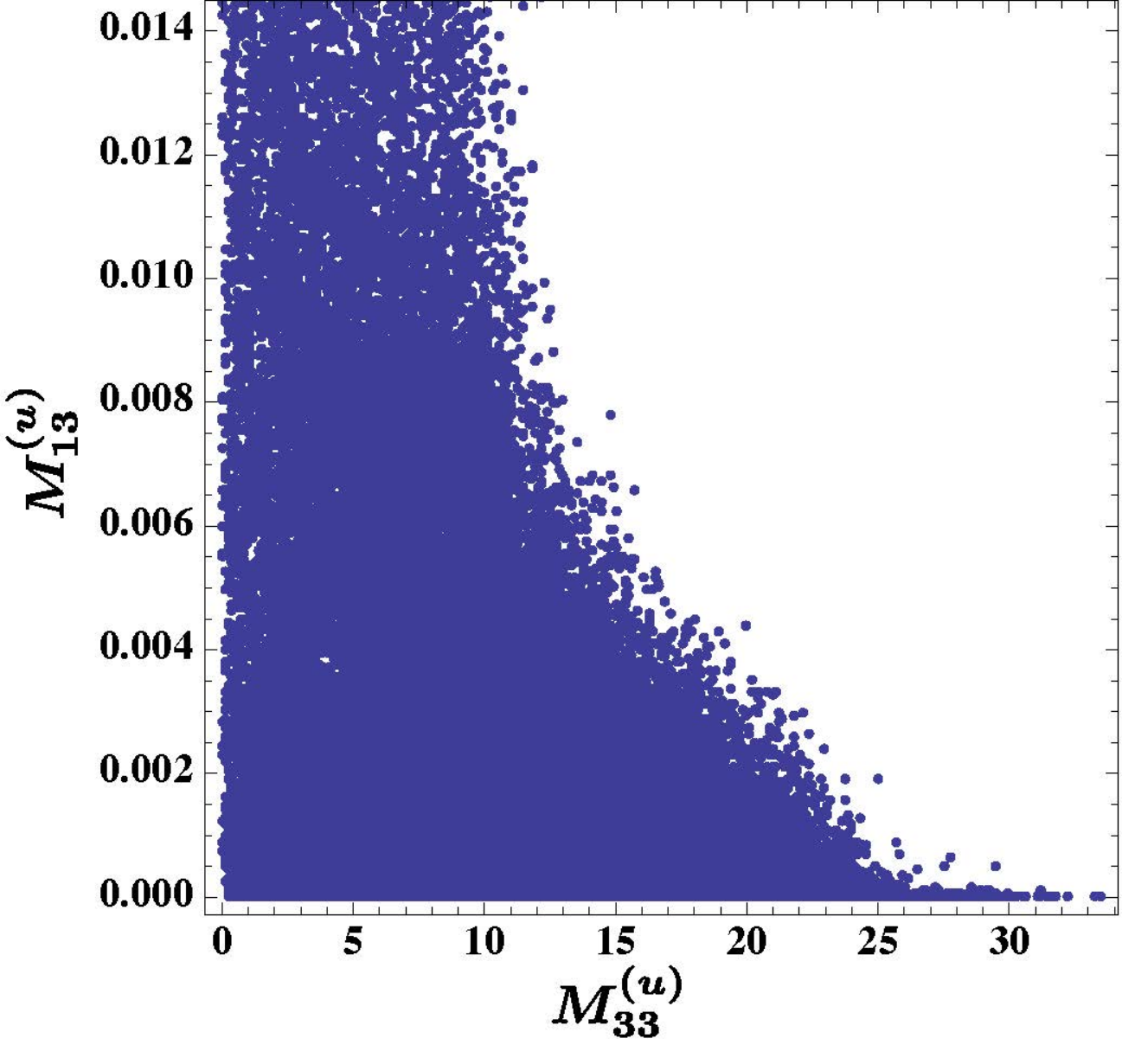}
\caption{
{
{The left} scatter plot shows the distribution of $M_{33}^{(u)} - M_{13}^{(u)}$ when we choose 100,000 points randomly around the configuration
in Eq.(\ref{a_setofsolution}) within $\pm 10\%$ being consistent with the order in Eq.~(\ref{positionorder1}).
{The right} one represents the same thing when we pick up 100,000 points randomly only with following the order in Eq.~(\ref{positionorder1}).
}
}
\label{randomprofile1_pdf}
\end{figure}

%\begin{figure}[t]
%\centering
%\includegraphics[width=0.35\columnwidth]{A4_Mu23-Mu22-master.pdf}
%\includegraphics[width=0.36\columnwidth]{A4_Mu13-Mu33-master.pdf}
%\caption{
%{
%Left (right) scatter plot shows the distribution of $M_{22}^{(u)} - M_{23}^{(u)}$ ($M_{33}^{(u)} - M_{13}^{(u)}$) when we pick up 100,000 points randomly only with following the order in Eq.~(\ref{positionorder1}).}
%}
%\label{randomprofile2_pdf}
%\end{figure}

\begin{figure}[t]
\centering
\includegraphics[width=0.30\columnwidth]{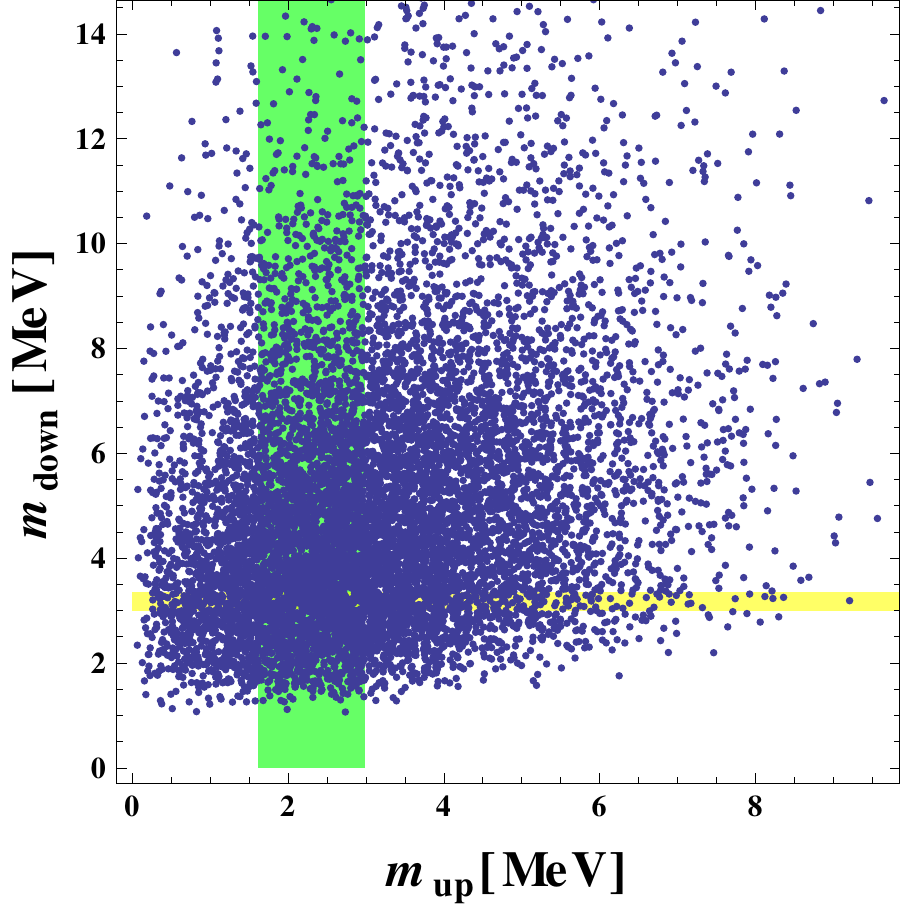}
\includegraphics[width=0.32\columnwidth]{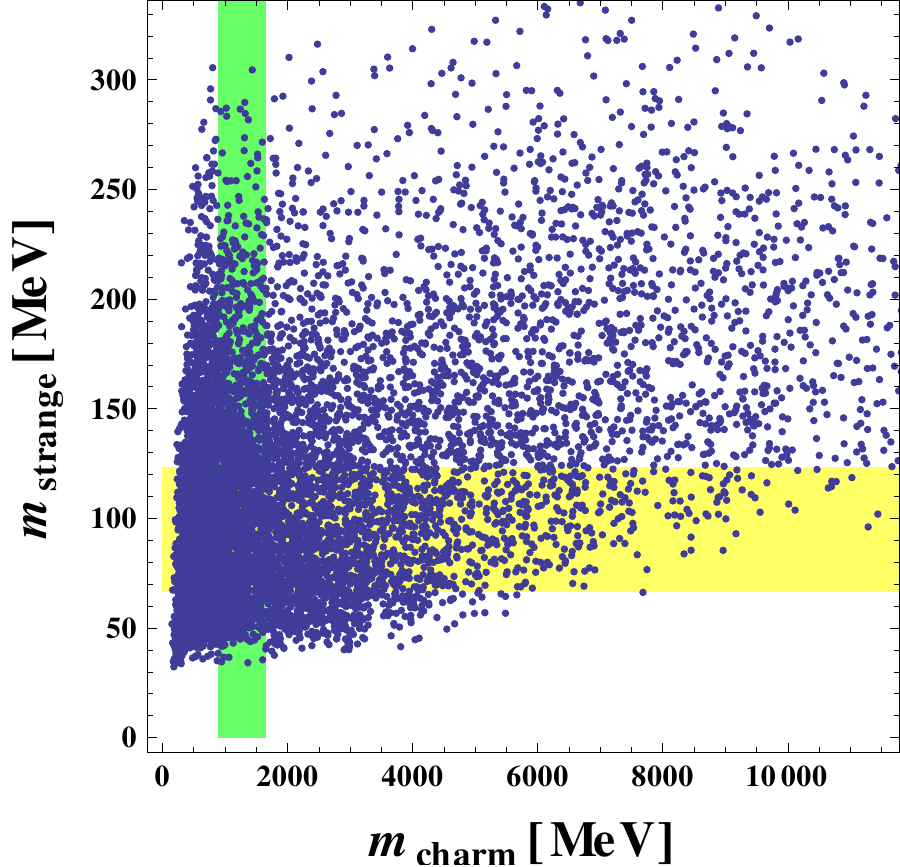}
\includegraphics[width=0.307\columnwidth]{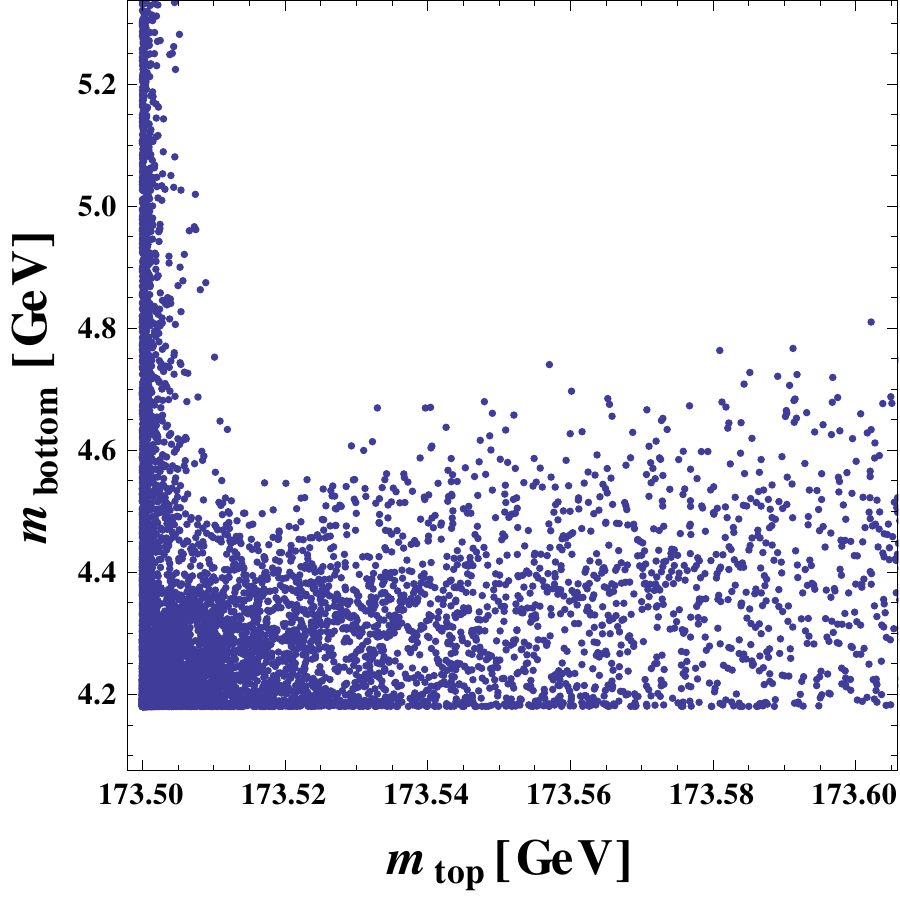}
\caption{
{
From left to right, distributions of $m_{\text{up}} - m_{\text{down}}$, $m_{\text{charm}} - m_{\text{strange}}$ and $m_{\text{top}} - m_{\text{bottom}}$ with 10,000 random points within $\pm 20\%$ parameter deviations from the central values in Eq.~(\ref{a_setofsolution} following the order in Eq.~(\ref{positionorder1}).
The green (yellow) band in the left and {center} plots represents $\pm 30\%$ range from the central values in Eq.~(\ref{a_setofsolution}).
In the right plot, we skip {depicting} the bands because all the shown ranges are covered by them.}
}
\label{distribution1_pdf}
\end{figure}

\begin{figure}[t]
\centering
\includegraphics[width=0.30\columnwidth]{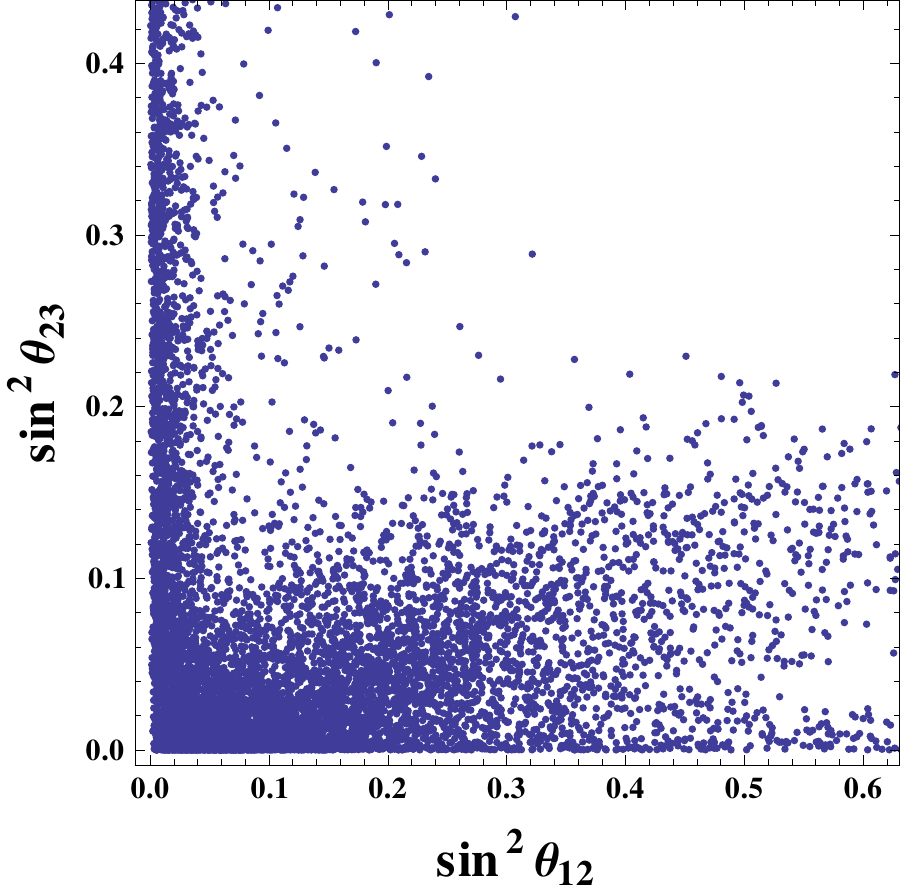}
\includegraphics[width=0.30\columnwidth]{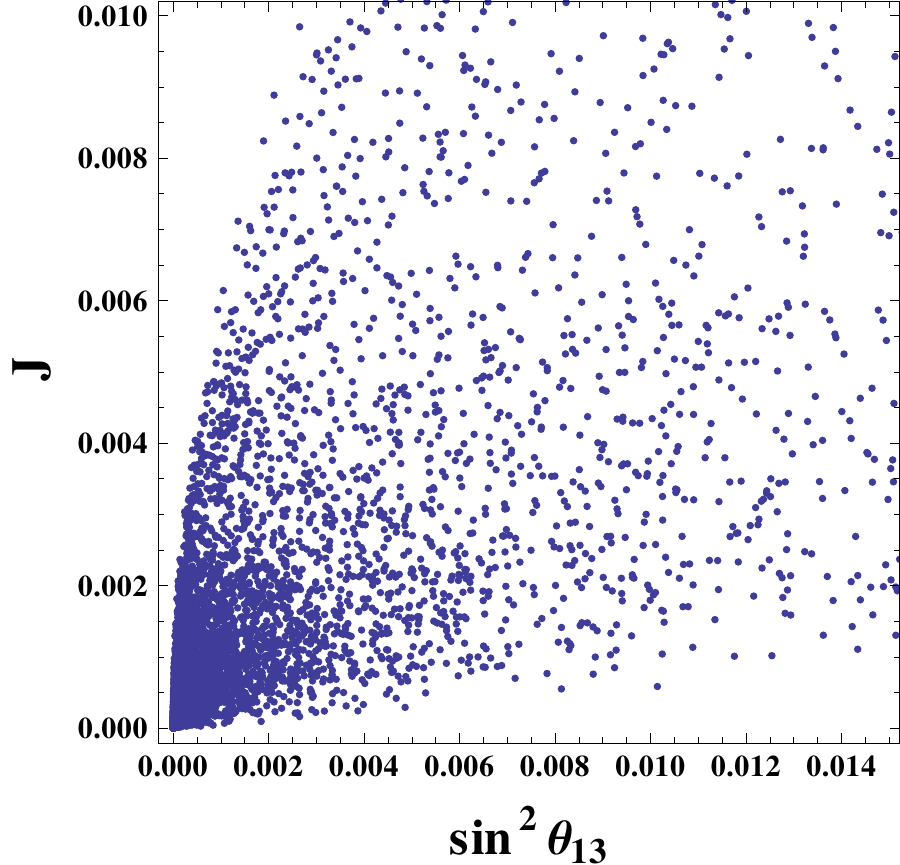}
\caption{
{
From left to right, distributions of $\sin^2{\theta_{12}} - \sin^2{\theta_{23}}$ and $\sin^2{\theta_{13}} - J$ with 10,000 random points within $\pm 20\%$ parameter deviations from the central values in Eq.~(\ref{a_setofsolution}) following the order in Eq.~(\ref{positionorder1}).
The allowed region is just near the point $(0,0)$ in both the plots.}
}
\label{distribution2_pdf}
\end{figure}

In this appendix, we discuss distribution patterns of quark mass-matrix elements and required orders in tuning the input parameters with the results given in {Sec.}~\ref{results} and \ref{results_CP}.

We first focus on the matrices in Eq.~(\ref{massmatrix1}).
In our model, the geometry of {the} extra dimension strongly restricts the form of the matrices.
In fact, we cannot fill all the elements of the mass matrices and at least three of the nine elements for each mass matrix have to be zero, as shown in Eq.~(\ref{massmatrix1}).
This property is contrasted with that of the {standard model}, where all the mass matrix elements are free parameters.
This fact means that possible patterns of mass matrices are constrained by the shape of the geometry of our model.

Furthermore, it turns out that the values of the {nonzero} elements in the mass matrices~(\ref{massmatrix1}) cannot be controlled freely.
To see this, we investigated correlations of matrix elements.
In the left figure of Fig.~\ref{randomprofile1_pdf}, we chose 100,000 points randomly around the configuration in Eq.(\ref{a_setofsolution}) within $\pm 10\%$ being consistent with the order in Eq.~(\ref{positionorder1}) and depicted the resultant values as two scatter plots ($M_{33}^{(u)} - M_{13}^{(u)}$).
The right figure of Fig.~\ref{randomprofile1_pdf} shows the same thing when we pick up 100,000 points randomly only with following the order in Eq.~(\ref{positionorder1}).

It follows from Fig.~\ref{randomprofile1_pdf} that we find no random distribution in the
$M^{(u)}_{33} - M^{(u)}_{13}$ plane and a strong correlation between $M^{(u)}_{13}$ and
$M^{(u)}_{33}$.
We further see the property that the typical value of $M^{(u)}_{13}$ is much smaller than
that of $M^{(u)}_{33}$.
In the quark sector, the mixing angles are known to be small, so that off-diagonal elements of the mass matrices will be preferred to be {subleading} compared with the diagonal ones with suitable magnitudes.
Our geometry realizes this point naturally via its geometry.

From the above observations, we may conclude that the quark mass matrices in our model are considerably restricted from the geometry of the extra dimension, and hence that it is {nontrivial} to reproduce the quark-related properties of the {standard model}, although we have 16 input parameters for quark profiles, where 13 parameters are independent, to explain the 10 {standard model} parameters (6 quarks masses and 4 CKM parameters).

As another consideration, we investigate the resultant quark masses and CKM matrix elements when we change the input parameters around the central values in Eq.~(\ref{a_setofsolution}) and then find that the quark masses and CKM matrix elements are very sensitive to some of the input parameters.

To show the sensitivity of the input parameters, we first alter all the parameters randomly within $10\%$, $1\%$, {$0.1\%$,} and $0.01\%$, respectively, obeying the order in Eq.~(\ref{positionorder1}){,} and calculate the masses and the elements.
When we proceed with the above procedure 100,000 times in each case,
11, 11,039, 81,955{,} and 100,000 points survive after putting the cut where all the resultants are within $15\%$.
These results indicate that parameter tuning less than $1\%$ is required as our inputs are so in Eq.~(\ref{a_setofsolution}).
The scatter plots in Figs.~\ref{distribution1_pdf} and \ref{distribution2_pdf} represent the distributions of the physical parameters with 10,000 random points within $\pm 20\%$ around the central values, where the CKM angles and the Jarlskog parameter are apt to getting away from the required range easily, {for which the} experimental central values are $\sin^2{\theta_{12}} \sim 0.05$, $\sin^2{\theta_{23}} \sim 0.002$, $\sin^2{\theta_{13}} \sim 0.00001${,} and $J \sim 0.00003$, respectively.
This issue is explained by the fact that off-diagonal elements of the CKM matrix are closely related to those of the up- and down-quark mass matrices, which at least parts of them are, sensitive to perturbation of the input parameters. 
We can also find the tendency that bottom and top masses are not away from the central values by the perturbation.

On the other hand, we try to alter an input parameter separately.
In each case, points of the numbers in {Table}~\ref{table:individualchange} pass the cut 
which rejects the possibilities that at least one resultant value is out of the $\pm 15\%$ deviation range from the central value in Eq.~(\ref{a_setofsolution}). 
According to the result, we understand that quark masses and mixing angles are sensitive to the positions of point interactions, while those are insensitive to (absolute values of) the bulk masses and the twisted angle.
Then we can conclude that parameter tuning in $M_Q$, $M_{\mathcal{U}}$, $M_{\mathcal{D}}${,} and $\theta$ in Eq.~(\ref{a_setofsolution}) is not always necessary.

Finally, we briefly comment on the required orders of significant digits in the input parameters.
As we have discussed before based on {Table}~\ref{table:individualchange}, the system is insensitive to $M_Q$, $M_{\mathcal{U}}$, $M_{\mathcal{D}}${,} and $\theta$ around the central region of the parameters{,} and then single digits are sufficient for them.
On the other hand{,} for $L_{2}^{(q)}$ and $L_{2}^{(d)}$, as also expressed in {Table}~\ref{table:individualchange}, triple digits are required because of their great sensitivity.
For the other values, tuning {up to} double digits is enough for our purpose since they are less sensitive than $L_{2}^{(q)}$ and $L_{2}^{(d)}$ as shown in {Table}~\ref{table:individualchange}.

\begin{table}[t]
\begin{center}
\begin{tabular}{|c||c|c|c|c|c|c|c|c|c|c|c|c|}
\hline
parameter &
$L_{1}^{(q)}$ &
$L_{2}^{(q)}$ &
$L_{0}^{(u)}$ &
$L_{1}^{(u)}$ &
$L_{2}^{(u)}$ &
$L_{0}^{(d)}$ &
$L_{1}^{(d)}$ &
$L_{2}^{(d)}$ &
$M_Q$ &
$M_{\mathcal{U}}$ &
$M_{\mathcal{D}}$ &
$\theta$
\\ \hline \hline
$\#$ of surviving points&
105 &
11 &
46 &
85 &
42 &
270 &
172 &
10 &
679 &
730 &
372 &
971
\\ \hline
\end{tabular}
\caption{
{
Numbers of surviving points out of 1,000 ones after the cut where all the physical resultants are within $\pm 15\%$ with changing {an} input parameter within $\pm 10\%$  from the central values in Eq.~(\ref{a_setofsolution}) individually.
}
}
\label{table:individualchange}
\end{center}
\end{table}

%\bibliographystyle{TitleAndArxiv}
%\bibliographystyle{JHEP}
%\bibliography{maintextref}

\bibliographystyle{utphys}
\bibliography{sakamoto_letter_CP}

%\bibliographystyle{TitleAndArxiv}
%\bibliography{letter}

\end{document}